\documentclass[prd,12pt,superscriptaddress]{revtex4}%
\usepackage{amsmath}
\usepackage{amsfonts}
\usepackage{amssymb}
\usepackage{graphicx}%
\usepackage{amsmath,amssymb}
\usepackage{mathrsfs}
\usepackage{graphicx}
\usepackage{color}
\usepackage{subfigure}
\usepackage{fancyhdr}
\usepackage{multirow}
\usepackage{float}
\usepackage{epsfig}
\usepackage{amsfonts}
\usepackage{bm}

\def\be{\begin{equation}}
\def\ee{\end{equation}}
\def\ba{\begin{eqnarray}}
\def\ea{\end{eqnarray}}

\begin{document}

\title{Kiselev-inspired Wormholes}

\author{Jureeporn Yuennan} 
\email{jureeporn_yue@nstru.ac.th}
\affiliation{Faculty of Science and Technology, Nakhon Si Thammarat Rajabhat University, Nakhon Si Thammarat, 80280, Thailand}

\author{Piyachat Panyasiripan}
\email{tampiya.5545@gmail.com}
\affiliation{The Institute for Fundamental Study “The Tah Poe Academia Institute”, Naresuan University, Phitsanulok 65000, Thailand}

\author{Phongpichit Channuie}
\email{phongpichit.ch@mail.wu.ac.th}
\affiliation{School of Science, Walailak University, Nakhon Si Thammarat, 80160, Thailand}
\affiliation{College of Graduate Studies, Walailak University, Nakhon Si Thammarat, 80160, Thailand}

\date{\today}

\begin{abstract}
In this study, we investigate traversable wormholes inspired by the Kiselev framework, which extends classical black hole solutions by incorporating anisotropic fluids. These exotic fluids play a crucial role in cosmology, particularly in explaining phenomena such as the accelerated expansion of the universe. We generalize the Kiselev framework to static, spherically symmetric traversable wormholes and analyze their properties under two distinct models of the redshift function: a constant redshift function and one that varies inversely with the radial coordinate. We examine the energy conditions—specifically the Null Energy Condition (NEC), Weak Energy Condition (WEC), and Strong Energy Condition (SEC)—for these models, demonstrating that only certain exotic fluids can sustain the wormhole structure. Furthermore, we quantify the amount of exotic matter required to maintain these wormholes using the volume integral quantifier and compare our results with other wormhole models. Additionally, we compute the effective potential for photons in Kiselev-inspired wormholes under both redshift function models and analyze their implications for weak gravitational lensing. Our findings suggest that Kiselev-inspired wormholes could serve as viable candidates for exotic geometries, potentially paving the way for future observational verification.

\end{abstract}

\maketitle

\section{Introduction}

The Null Energy Condition (NEC) states that for any null vector, 
$T_{\mu\nu}k^{\mu}k^{\nu} \geq 0$, where $T_{\mu\nu}$ represents the Stress-Energy Tensor (SET). In terms of energy density $\rho$ and pressures $p_i$, the NEC is expressed as:
\[
\rho + p_i \geq 0 \quad \text{for} \quad i = 1, 2, 3.
\]
A traversable wormhole (TW) is a solution of the Einstein Field Equations (EFE) that connects two distant regions through a tunnel, which violates the NEC. Specifically, if $p_r$ is the radial pressure, the following inequality must be satisfied:
\[
\rho + p_r \leq 0.
\]
The initial concept of a traversable wormhole was introduced by Ludwig Flamm \cite{Flamm:1916} and later expanded by Einstein and Rosen \cite{Einstein:1935tc}, who formulated the Schwarzschild solution to describe a bridge-like structure known as the Einstein-Rosen (ER) bridge. However, it was not until the work of Morris and Thorne \cite{MT, Morris:1988tu} that the modern framework for TWs was developed \cite{Visser1995}. Interestingly, the Schwarzschild solution describes a wormhole that is not traversable. The violation of the NEC is associated with the presence of "exotic" matter, suggesting that a semiclassical or quantum theory of gravity may be essential for a deeper understanding of TWs. Casimir energy, a form of vacuum energy confirmed experimentally, is a likely candidate for stabilizing a TW. The Generalized Uncertainty Principle (GUP) \cite{Jusufi:2020rpw}, which includes electric charge considerations \cite{Samart:2021tvl, Garattini:2023qyo}, extends this idea. The Casimir effect has also been explored in relation to TWs within the context of modified gravity theories, as discussed in \cite{EPJC1, Garattini:2020kqb, Garattini:2021kca, Santos:2021jjs}. Casimir wormholes have been studied in this framework, with examples found in the literature on modified gravity theories \cite{Hassan:2022hcb, Sokoliuk:2022jcq, Oliveira:2021ypz}.

Models that include a component with an arbitrary equation of state $\omega = p/\rho$ are known as dark energy models. For cosmic acceleration, $\omega < -1/3$ is necessary, while $\omega < -1$ enters the realm of phantom energy. Although numerous theories and models have been proposed to explain dark energy, a satisfactory understanding remains elusive. The Kiselev framework offers a powerful extension to classical black hole solutions by incorporating anisotropic fluids such as quintessence and phantom energy. These exotic forms of matter are linked to phenomena like the accelerated expansion of the universe, making them highly relevant for modern cosmology and gravitational theory. The general form of the line element for a spherically symmetric Kiselev black hole is \cite{Kiselev:2002dx}
\ba
    ds^2 = -\left(1 - \frac{2M}{r} - \frac{K}{r^{3\omega + 1}}\right) dt^2 + \left(1 - \frac{2M}{r} - \frac{K}{r^{3\omega + 1}}\right)^{-1} dr^2 + r^2 d\Omega^2,
\ea
where \( M \) is the black hole mass, \( K \) is a contant, \( \omega \) is the equation of state parameter of the fluid, and \( d\Omega^2 \) is the metric of a 2-sphere. This formulation alters the black hole’s horizon structure and thermodynamics, offering new insights into how exotic fluids influence gravitational systems. One reason for the popularity of this model lies in its versatility: it reduces to the Schwarzschild solution when \(\omega = 0\), the Reissner--Nordstr\"om solution when \(\omega = 1/3\), and the Schwarzschild-(anti)-de Sitter solution when \(\omega = -1\). It was asserted to the terminology confusion by \cite{Visser:2019brz} that the Kiselev black hole is neither perfect fluid, nor is it quintessence.  However, while black holes are primarily associated with horizons and singularities, wormholes represent a different class of solutions: hypothetical structures that can connect distant regions of spacetime and potentially allow for traversable paths. Therefore, expanding the Kiselev framework from black holes to wormholes is an engaging and promising line of investigation.

This work is structured as follows. In Section (\ref{KW}), we introduce the Kiselev wormhole metric and derive the equations governing its geometry and matter content. Section (\ref{em}) presents the embedding diagram, providing a visualization of the shape of the wormhole. In Section (\ref{EC}), we analyze the energy conditions for two models of the redshift function, $\Phi=const$. and $\Phi=r_{0}/r$, exploring their implications for the wormhole's traversability and stability. Section (\ref{AM}) discusses the amount of exotic matter required to sustain the wormhole using the volume integral quantifier. We also ompute the effective potential for photons in Kiselev-inspired wormholes, considering two distinct redshift functions: constant and inversely dependent on the radial coordinate in Section (\ref{secvi}). In Section (\ref{len}), we explore the effects of weak gravitational lensing in these wormhole configurations. Finally, we present our conclusions in the last section.

\section{Kiselev wormholes}\label{KW}
We extend the Kiselev framework to wormholes by exploring how similar anisotropic matter distributions can support traversable wormholes. In the case of a static, spherically symmetric traversable wormhole, the metric takes the general form
\ba
ds^2 = -e^{2\Phi(r)} dt^2 + \left(1 - \frac{b(r)}{r}\right)^{-1}dr^2 + r^2 d\Omega^2,\label{ds}
\ea
where \( \Phi(r) \) is the redshift function, ensuring no event horizons, and \( b(r) \) is the shape function, which defines the wormhole’s geometry. Furthermore, the shape function \( b(r) \) defines the geometry of the wormhole, specifically through the condition \( b(r_{0}) = r_{0} \), where \( r_{0} \) represents the radius of the wormhole throat. As a result, the shape function must also satisfy the flaring-out condition \cite{MT}:
\begin{equation}
    \frac{b(r) - r b^{\prime}(r)}{b^{2}(r)} > 0,
\end{equation}
which requires that \( b^{\prime}(r_{0}) < 1 \) at the throat of the wormhole. With
the help of the line element $\left(\ref{ds}\right)  $, we obtain the
following set of equations resulting from the energy-momentum components to
yield ($G=1$):
\begin{equation}
\frac{b^{\prime}(r)}{r^{2}}=8\pi\rho(r),\label{rho}%
\end{equation}%
\begin{equation}
\left[  2\left(  1-\frac{b(r)}{r}\right)  \frac{\Phi^{\prime}(r)}{r}%
-\frac{b(r)}{r^{3}}\right]  =8\pi p_{r}(r)\label{pr}%
\end{equation}
and%
\begin{align}
&  \left(  1-\frac{b\left(  r\right)  }{r}\right)  \left[  \Phi
^{\prime\prime}(r)+\Phi^{\prime}(r)\left(  \Phi^{\prime}(r)+\frac{1}%
{r}\right)  \right] -\frac{b^{\prime}\left(  r\right)  r-b\left(  r\right)  }{2r^{2}}\left(
\Phi^{\prime}(r)+\frac{1}{r}\right) =8\pi p_{t}(r),\label{pt}%
\end{align}
where $p_{t}=p_{\theta}=p_{\phi}$. In the Kiselev-inspired wormhole, we model the surrounding energy-momentum tensor \( T_{\mu\nu} \) using an anisotropic fluid distribution. The anisotropic stress arises from components such as quintessence or phantom energy, with the following state-dependent energy density \cite{Visser:2019brz}:
\begin{equation}
    \rho(r) = \frac{c}{8\pi r^{3(\omega +1)}}.
\end{equation}
The corresponding radial and tangential pressures, \( p_r(r) \) and \( p_t(r) \), can be straightforwardly obtained from Eq.(\ref{pr}) and Eq.(\ref{pt}), respectively. Using the above energy density, the shape function can be directly determined to obtain
\ba
b(r)=-\frac{c\,r^{-3 \omega}}{3 \omega}+c_1\,.
\ea
We use $b(r_{0}) = b_{0} = r_{0}$ and $c_{1}=0$, and then calculate the constant $c$ to obtain $c=-3 \omega\,r_{0}^{3 \omega +1}$. Therefore, the shape function takes the form
\ba
b(r)=r_{0}\,\Big(\frac{r_{0}}{r}\Big)^{3\omega}\,.\label{br}
\ea
Notice that the solution (\ref{br}) for $\omega>0$ is asymptotically flat, meaning that as $r\rightarrow \infty$, i.e., the spacetime becomes increasingly flat. On the other hand, when $\omega<0$, the solution deviates from asymptotic flatness, indicating that the spacetime remains curved even at large distances from the wormhole throat. We also see that $b(r)/r=0$ for $\omega>-1/3$. With the flaring-out condition \( b^{\prime}(r_{0}) < 1 \), we find that 
\ba
b'(r)|_{r=r_{0}}=-3 \omega  \left(\frac{r_{0}}{r}\right)^{3 \omega +1}|_{r=r_{0}} <1\,. \label{omega}
\ea
Therefore, the parameter $\omega$ is constrained to be \(\omega > - \tfrac{1}{3}\). This constraint on $\omega$ suggests that only certain exotic fluids can maintain the wormhole structure. Specifically, the matter distribution should be less exotic than phantom energy, which typically has $\omega<-1$. Fluids with $-1/3<\omega<0$ are good candidates and do not severely violate energy conditions.

\subsection{Model $\Phi=const.$}
Introducing the scaling of coordinate $\exp(2\Phi)\,dt^{2} \rightarrow dt^{2}$ (since $\exp(2\Phi) = const.$), the wormhole metric reads
\ba
ds^2 &=& -dt^2 + \frac{1}{1 - \Big(\frac{r_{0}}{r}\Big)^{1+3\omega}}dr^2 \nonumber\\&+& r^2 d\Omega^2. \label{ds1}
\ea
With the redshift function $\Phi=const.$, this wormhole avoids event horizons, ensuring that travelers can pass through without encountering regions of infinite time dilation. This choice is widely used in the literature as it simplifies the analysis while still yielding physically meaningful solutions. Setting $\Phi=0$ eliminates any gravitational redshift effects, which simplifies the metric and the corresponding field equations. More importantly, it ensures that there are no event horizons, maintaining the traversability of the wormhole. This assumption has been frequently employed in various works on traversable wormholes, see Refs.\cite{MT,Visser1995}.

\subsection{Model $\Phi=r_{0}/r$}
In this case, the wormhole metric takes the form
\ba
ds^2 = -e^{2r_{0}/r}dt^2 + \frac{1}{1 - \Big(\frac{r_{0}}{r}\Big)^{1+3\omega}}dr^2 + r^2 d\Omega^2. \label{ds1}
\ea
This form of the redshift function represents a non-trivial gravitational potential that remains finite everywhere, including at the throat ($r=r_{0}$). This choice ensures that the metric coefficient $g_{tt}$ does not vanish, thereby avoiding the formation of horizons and allowing smooth traversal through the wormhole. Additionally, this functional form is inspired by previous studies that explored wormholes supported by various exotic matter distributions and analyzed their stability under perturbations Refs.\cite{Lobo:2005us,Jusufi:2020rpw}. The $\Phi=r_{0}/r$ function introduces a redshift effect that gradually diminishes at large distances, making it relevant in astrophysical contexts where the gravitational potential decays with radial distance. This configuration warrants further discussion regarding the implications for the geometry, traversability, and energy conditions. As $r\rightarrow \infty$, $\Phi\rightarrow 0$, meaning the redshift effect diminishes at large distances, which aligns with asymptotic flatness when $\omega>0$. 

\section{Embedding diagram}\label{em}

In this section, we visualize the wormhole solution using an embedding diagram. To achieve this, the analysis focuses on an equatorial slice (\(\theta = \pi/2\)) at a fixed moment in time (\(t = {\rm const.}\)). In this configuration, the wormhole metric is written as
\begin{equation}
    ds^2 = \frac{dr^2}{1 - \frac{b(r)}{r}} + r^2 d\phi^2,
\end{equation}
where \(b(r)\) is the shape function. To visualize the wormhole, this metric is embedded in three-dimensional Euclidean space, represented in cylindrical coordinates:
\begin{equation}
    ds^2 = dz^2 + dr^2 + r^2 d\phi^2.
\end{equation}

\begin{figure}[ht!]
    \centering
\includegraphics[width=3in,height=3in,keepaspectratio=true]{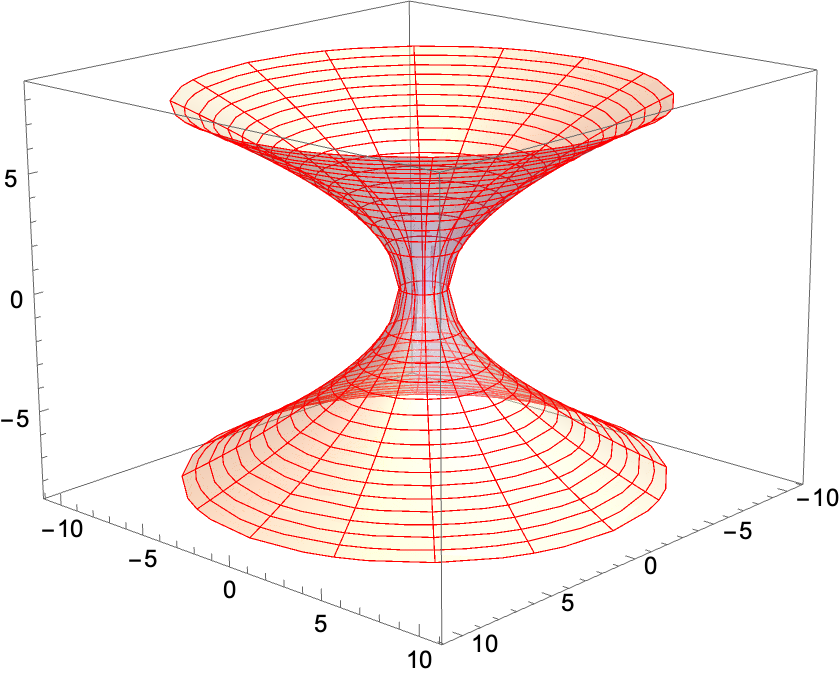}
\includegraphics[width=3in,height=3in,keepaspectratio=true]{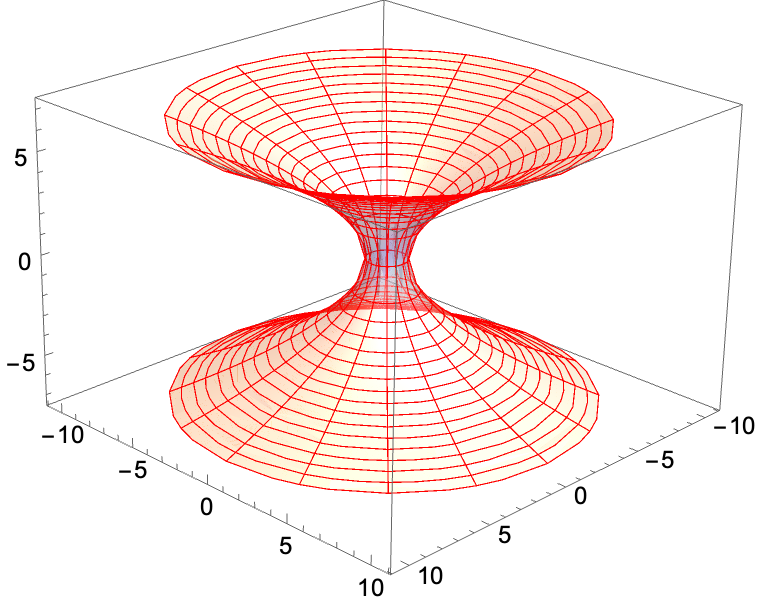}
    \caption{The plots show the wormhole embedded in a three-dimensional Euclidean space. Upper panel: we have used $r_{0} = 1$ and $\omega=-1/6$. Lower panel: we have used $r_{0} = 1$ and $\omega=-1/20$.}
    \label{emb}
\end{figure}

By comparing these two metrics, the following relation for the wormhole’s shape emerges:
\begin{equation}
    \frac{dz}{dr} = \pm \sqrt{\frac{r}{r - b(r)} - 1}.
\end{equation}
Here, \(b(r)\) is the shape function from Eq.~(\ref{br}). The results are illustrated in Fig.~\ref{emb}, which shows how the wormhole’s shape changes based on different values of the $\omega$ parameter.

\section{Energy Conditions}\label{EC}
\subsection{Model $\Phi=const.$}
The stress-energy tensor components are derived from the Einstein field equations and are given by:
\ba
\rho(r) &=& \frac{c}{8\pi r^{3(\omega +1)}}=-\frac{3 \omega\,r_{0}}{8\pi r^3} \left(\frac{r_{0}}{r}\right)^{3m}\,,\\p_r(r) &=& -\frac{1}{8\pi}\frac{b(r)}{r^3}=-\frac{r_{0}}{8 \pi  r^3}\left(\frac{r_{0}}{r}\right)^{3 \omega }\\p_{t}(r)&=&\frac{r_{0} (3 \omega +1)}{16 \pi  r^3}\left(\frac{r_{0}}{r}\right)^{3 \omega }\,.
\ea
We first check for the Null Energy Condition (NEC). The NEC states that:
\(\rho + p_r \geq 0\). We find
\ba
\rho + p_r &=& -\frac{r_{0} (3 \omega +1)}{16 \pi  r^3}\left(\frac{r_{0}}{r}\right)^{3 \omega }\, \\
&=& -\frac{(1+3\omega)}{8\pi} \cdot \frac{b(r)}{r^3}\nonumber\\\rho+2p_{t}&=&\frac{r_{0} \left(\frac{r_{0}}{r}\right)^{3 \omega }}{8 \pi  r^3}\,.
\ea
Therefore the NEC is satisfied if \(3\omega + 1 < 0\), i.e., \(\omega < -\frac{1}{3}\) and violated if \(3\omega + 1 > 0\), i.e., \(\omega > -\frac{1}{3}\). Next we examine the Weak Energy Condition (WEC). It is states that \(\rho \geq 0 \quad \text{and} \quad \rho + p_r \geq 0\). Since \(\rho > 0\) for negative \(\omega\), the WEC is satisfied for most values of \(\omega < -\frac{1}{3}\) and violated if \(3\omega + 1 > 0\), i.e., \(\omega > -\frac{1}{3}\). Lastly, let us check for the Strong Energy Condition (SEC). The SEC states that \(\rho + 2p_t \geq 0\) and \(\rho + p_r + 2p_t \geq 0\). We find that \(\rho + 2p_t \geq 0\) and  \(\rho + p_r + 2p_t = 0\) for this model.
\begin{figure}[ht!]
    \centering
\includegraphics[width=3in,height=3in,keepaspectratio=true]{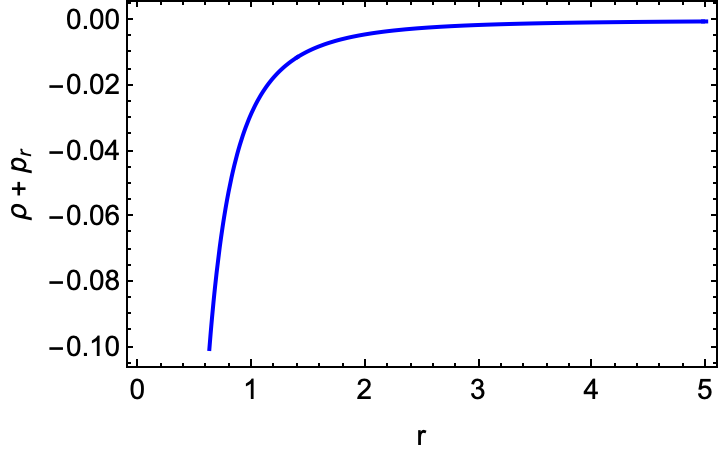}
\includegraphics[width=3in,height=3in,keepaspectratio=true]{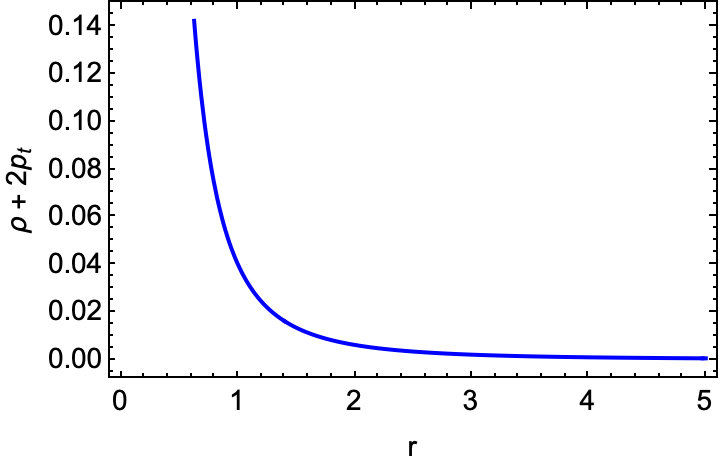}
    \caption{The plots show the variation of the energy conditions as a function of $r$ using $\Phi=const.$: $\rho + p_{r}$ (upper panel) and $\rho +2p_{t}$ (lower panel). We use $\omega=-1/10$.}
    \label{TUUT}
\end{figure}

For values \(\omega > -\frac{1}{3}\), the NEC is violated, indicating that exotic matter is required to maintain the wormhole. The constraint $\omega > -\frac{1}{3}$ for NEC violation means that the matter supporting the wormhole cannot be more exotic than phantom energy, which typically has $\omega < -\frac{1}{3}$. Fluids with $-\frac{1}{3}<\omega<0$ are ideal candidates for sustaining the wormhole without extreme violations of energy conditions. These results suggest that traversable wormholes in this model can be maintained with forms of matter that only mildly violate the known energy conditions, rather than requiring extreme or unphysical exotic matter.

\subsection{Model $\Phi=r_{0}/r$}
The stress-energy tensor components are derived from the Einstein field equations and are given by
\ba
\rho(r) &=& \frac{c}{8\pi r^{3(\omega +1)}}=-\frac{3 \omega\,r_{0}}{8\pi r^3} \left(\frac{r_{0}}{r}\right)^{3m}\,,\\p_r(r) &=& \frac{r_{0} (2 r_{0}-r) \left(\frac{r_{0}}{r}\right)^{3 \omega }-2 r r_{0}}{8 \pi  r^4}\,,\\p_{t}(r)&=&\frac{2 r r_{0} (r+r_{0})}{16 \pi  r^5}+\frac{r_{0} \left(\frac{r_{0}}{r}\right)^{3 \omega } \left(r^2 (3 \omega +1)-3 r r_{0} (\omega +1)-2 r_{0}^2\right)}{16 \pi  r^5}\,.
\ea
We first check for the Null Energy Condition (NEC). The NEC states that:
\(\rho + p_r \geq 0\). We find
\begin{align*}
\rho + p_r = -\frac{r_{0} (3 r \omega +r-2 r_{0}) \left(\frac{r_{0}}{r}\right)^{3 \omega }+2 r r_{0}}{8 \pi  r^4}\,.
\end{align*}
We find that for this model
\ba
\rho + p_r + 2p_t &=& \frac{r_{0}^2}{8 \pi  r^5}\Big(2 r-\left(\frac{r_{0}}{r}\right)^{3 \omega } (3 r \omega +r+2 r_{0})\Big)\,,\\\rho+2p_{t}&=&\frac{2 r r_{0} (r+r_{0})}{8 \pi  r^5}+\frac{r_{0} \left(r^2-3 r r_{0} (\omega +1)-2 r_{0}^2\right) \left(\frac{r_{0}}{r}\right)^{3 \omega }}{8 \pi  r^5}\,.
\ea
Therefore the NEC is violated for any values of \(\omega\), while the WEC is also violated since \(\rho + p_r<0\). Lastly, let us check for the Strong Energy Condition (SEC). We find for the SEC that \(\rho + 2p_t \geq 0\) and \(\rho + p_r + 2p_t < 0\). 
\begin{figure}[ht!]
    \centering
\includegraphics[width=3in,height=3in,keepaspectratio=true]{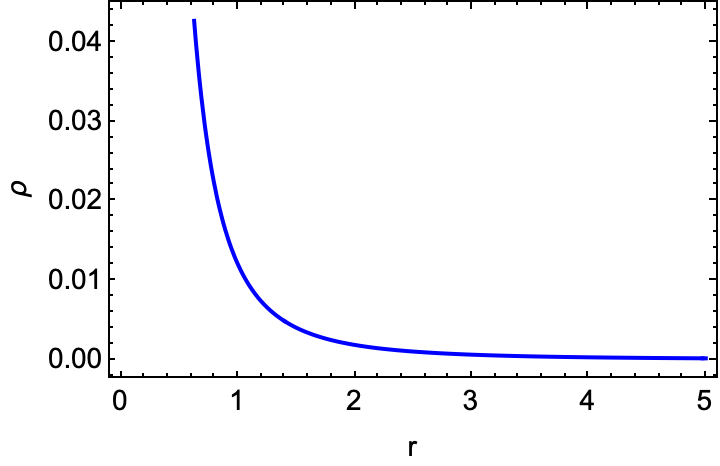}
\includegraphics[width=3in,height=3in,keepaspectratio=true]{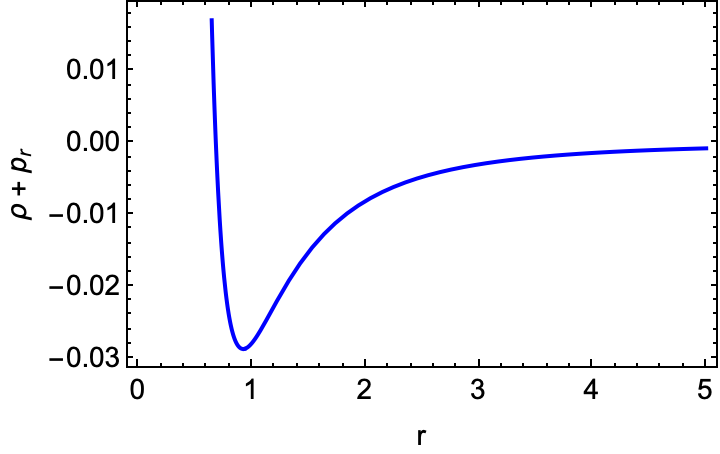}
\includegraphics[width=3in,height=3in,keepaspectratio=true]{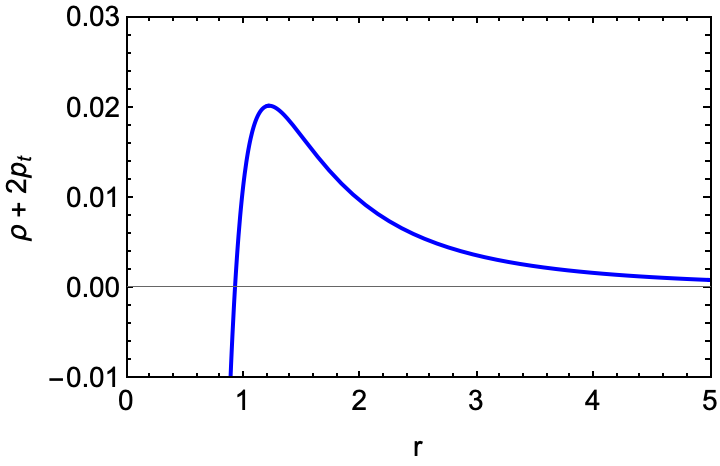}
\includegraphics[width=3in,height=3in,keepaspectratio=true]{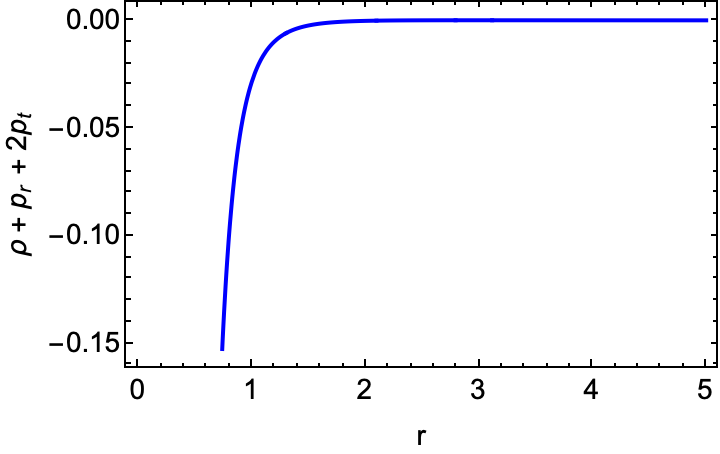}
    \caption{The plots show the variation of the energy conditions as a function of $r$ using $\Phi=r_{0}/r$: $\rho + p_{r}$ (upper panel) $\rho +2p_{t}$ (lower panel) and $\rho + p_{r} +2p_{t}$ (lowest panel). We use $\omega=-1/10$.}
    \label{TUUT}
\end{figure}

The introduction of the redshift function $\Phi=r_{0}/r$ modifies the energy conditions significantly compared to the constant $\Phi$ case.

\section{Amount of Exotic Matter}\label{AM}

In this section, we briefly discuss the “volume integral quantifier,” which measures the amount of exotic matter needed to sustain a wormhole. This quantifier is concerned solely with the quantities \( \rho \) and \( P_r \), and is defined by the following integral \cite{Visser:2003yf}
\ba
I_{V} = \oint [\rho + P_r] dV = 2 \int_{r_0}^{\infty} (\rho + P_r) dV,
\ea
which can be further simplified as
\ba
I_{V} = 8\pi \int_{r_0}^{\infty} (\rho + P_r) r^2 dr.
\ea
As previously mentioned, this volume integral provides insights into the total amount of exotic matter present in the spacetime. We will now evaluate this integral for the specific shape function \( b(r) \). To make this calculation tractable, we introduce a cutoff at some finite radius \( a \), so the wormhole extends from \( r_0 \) to \( a \). The integral then becomes:
\ba
I_{V} = 8\pi \int_{r_0}^{a} (\rho + P_r) r^2 dr.
\ea
For the specific case where \( \Phi = const. \), the wormhole can be sustained by arbitrarily small quantities of exotic matter. Evaluating the integral gives
\ba
I_{V} = r_{0} (\omega +1) \Big(1-\left(\frac{r_{0}}{a}\right)^{3 \omega }\Big),
\ea
indicating that for small values of $a$, the quantity of exotic matter diminishes, suggesting that the wormhole can persist with minimal exotic matter. This is an appealing feature from a physical standpoint because it suggests that Kiselev-inspired wormholes may not need large quantities of exotic matter to be stable. In this case, the shape function takes a relatively simple form, and the matter content satisfies the necessary conditions for maintaining the wormhole throat without requiring extreme deviations from normal matter properties. Moreover, for the second case where \( \Phi = r_{0}/r \), evaluating the integral gives
\ba
I_{V} = \frac{1}{3} r_{0} \left(\left(-\frac{6 r_{0}}{3 a \omega +a}+\frac{1}{\omega }+3\right) \left(\frac{r_{0}}{a}\right)^{3 \omega }+6 \log \left(\frac{r_{0}}{a}\right)-\frac{1}{\omega }+\frac{6}{3 \omega +1}-3\right),
\ea
suggesting that exotic matter is more concentrated around the throat of the wormhole when $\Phi = r_{0}/r$, although it can still be minimized as $a\rightarrow r$. The amount of exotic matter required in Kiselev-inspired wormholes can be compared to other traversable wormhole solutions, such as those based on the Morris-Thorne framework or phantom energy models, see, e.g., Refs.\cite{Visser:2003yf,Kar:2004hc,Lobo:2005us}.

\section{Effective Potential for Photons in Kiselev-inspired Wormholes}\label{secvi}
Consider the wormhole metric given by
\ba
ds^2 = -e^{2\Phi(r)} dt^2 + \frac{dr^2}{1 - \frac{b(r)}{r}} + r^2 (d\theta^2 + \sin^2\theta \, d\phi^2),
\ea
where \( b(r) = r_0 \left(\frac{r_0}{r}\right)^{3\omega} \). For photons, \( ds^2 = 0 \), and in the equatorial plane (\( \theta = \pi/2 \)), we write
\ba
-e^{2\Phi(r)} \dot{t}^2 + \frac{\dot{r}^2}{1 - \frac{b(r)}{r}} + r^2 \dot{\phi}^2 = 0.
\ea
In this situation, energy \( E = e^{2\Phi(r)} \dot{t} \) and angular momentum \( L = r^2 \dot{\phi} \) are the conserved quantities. The radial equation of motion becomes:
\ba
\dot{r}^2 = E^2 - \frac{L^2}{r^2 e^{2\Phi(r)}} \left(1 - \frac{b(r)}{r}\right).
\ea
As a result, the effective potential can be read from the above relation to obtain
\ba
V_{\text{eff}}(r) = \frac{L^2}{r^2 e^{2\Phi(r)}} \left(1 - \frac{b(r)}{r}\right).
\ea
We now compute the effective potential for two cases. Let us start with \( \Phi(r) = \text{const.} \) 
For \( e^{2\Phi(r)} = 1 \), we have
\ba
V_{\text{eff}}(r) = \frac{L^2}{r^2} \left(1 - \frac{r_0 \left(\frac{r_0}{r}\right)^{3\omega}}{r}\right).
\ea
and the second case, \( \Phi(r) = \frac{r_0}{r} \), we find
for \( e^{2\Phi(r)} = e^{2r_0/r} \):
\ba
V_{\text{eff}}(r) = \frac{L^2}{r^2 e^{2r_0/r}} \left(1 - \frac{r_0 \left(\frac{r_0}{r}\right)^{3\omega}}{r}\right).
\ea

Here we note that the effective potential governs the motion of the photons. Peaks in \( V_{\text{eff}}(r) \) correspond to the photon sphere, defined by:
    \[
    \frac{dV_{\text{eff}}}{dr} = 0, \quad \frac{d^2V_{\text{eff}}}{dr^2} < 0.
    \]

\begin{figure}[ht!]
    \centering
\includegraphics[width=5in,height=5in,keepaspectratio=true]{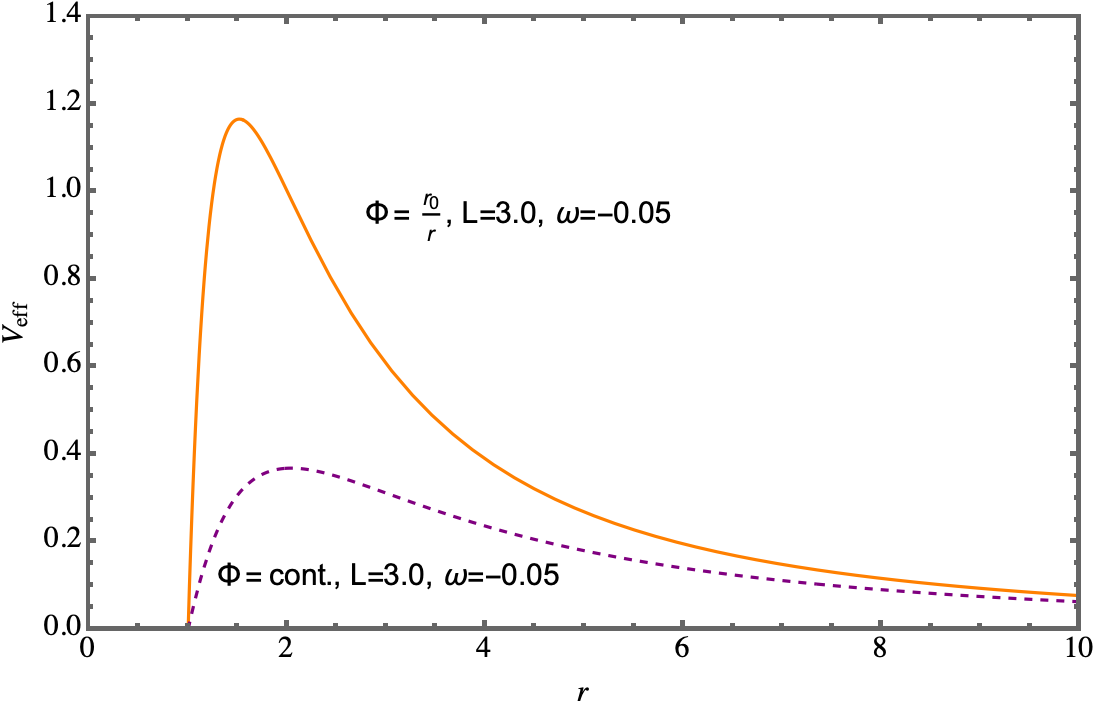}
    \caption{The plots show the radial profiles of the effective potential $V_{\rm eff}$ for timelike particles orbiting a non-rotating wormhole. We take $\Phi=r_{0}/r$ for a solid line and $\Phi={\rm cont.}$ for a dashed one, using $\omega =-0.05$ and $r_{0}=1$.}
    \label{Veff}
\end{figure}

The shape of \( V_{\text{eff}}(r) \) depends on the parameters \( r_0 \), \( \omega \), and the redshift function. The photon sphere, corresponding to a peak in the effective potential. We find from Fig.\ref{Veff} that the difference between constant- and inverse-redshift models, with the latter introducing additional curvature effects.

\section{Weak gravitational lensing}\label{len}
Gravitational lensing serves as a powerful method for exploring the spacetime geometry of wormholes by examining their gravitational influence on the trajectories of light. This technique has been utilized by many researchers to investigate various exotic wormhole configurations, as demonstrated by studies such as those cited in Refs. \cite{Jusufi:2017mav,Ovgun:2018tua,Ovgun:2018fnk,Nandi:2006ds,Ovgun:2018xys,Abe:2010ap,Kuhfittig:2013hva}. Through the analysis of gravitational lensing effects, we have gained valuable insights into the characteristics and behavior of these exotic wormholes, providing information on their stability, geometry, and implications within the framework of general relativity. Moreover, if gravitational lensing observations reveal unusual mass distributions or light bending patterns that cannot be accounted for by conventional matter, this could suggest the existence of exotic matter potentially linked to wormholes.

The Gauss-Bonnet theorem provides a geometric way to determine the weak gravitational lensing or deflection angle by integrating the Gaussian curvature over a surface, typically relevant in scenarios such as black holes and wormholes. Different authors utilize this technique to calculate the deflection angle of light by various black holes and wormholes spacetime, see, e.g., \cite{Gibbons:2008hb,Gibbons:2008zi,Gibbons:2011rh,Bloomer:2011rd,Werner:2012rc,Gibbons:2015qja,Ishihara:2016vdc,Das:2016opi,Sakalli:2017ewb,Jusufi:2017lsl}. In this section, we closely follow the works done by Refs.\cite{Gibbons:2008rj}. We write the metric for a wormhole in spherically symmetric spacetime as follows:
\begin{equation}
    ds^2 = -A(r) \, dt^2 + \frac{dr^2}{B(r)} + r^2 d\Omega^2,
\end{equation}
where the metric functions \(A(r)\) and \(B(r)\) are defined as:
\begin{equation}
    A(r) = \exp(2\Phi), \quad B(r) = 1 - \frac{b(r)}{r},\label{ab}
\end{equation}
and \(d\Omega^2 = d\theta^2 + \sin^2\theta \, d\phi^2\). In the equatorial plane \((\theta = \pi/2)\), the optical metric is derived by setting \(ds^2 = 0\):
\begin{equation}
    dt^2 = \frac{dr^2}{A(r) B(r)} + \frac{r^2 d\phi^2}{A(r)}.
\end{equation}
The non-zero Christoffel symbols read \cite{Kumaran:2021rgj,Javed:2020mjb}
\ba
    \Gamma^0_{00} &=& -\frac{B'(r)}{2B(r)} - \frac{A'(r)}{2A(r)}, \\ 
    \Gamma^1_{01} &=& \frac{1}{r} - \frac{A'(r)}{2A(r)},\\
    \Gamma^0_{11} &=& -rB(r)+\frac{r^{2}A'(r)B(r)}{2A(r)}\,,
\ea
where \(A'(r)\) and \(B'(r)\) are the derivatives with respect to \(r\). The Ricci scalar \(R\) for the optical metric is
\ba
    R = -\frac{A(r) B'(r)}{r} + \frac{A'(r) B'(r)}{2} + \frac{A'(r) B(r)}{r} + A''(r) B(r) - \frac{(A'(r))^2 B(r)}{2A(r)}.\label{RR}
\ea
The Gaussian curvature \(K\) is related to the Ricci scalar by
\begin{equation}
    K = \frac{R}{2}.
\end{equation}
The deflection angle of light is calculated using the Gauss-Bonnet theorem (GBT) \cite{Gibbons:2008rj}
\begin{equation}
    \alpha = - \int_0^{\pi} \int_{u/\sin \phi}^{\infty} K \, \sqrt{\det(g)} \, dr \, d\phi,
\end{equation}
where the term $\det(g)$ accounts for the area element of the optical metric, ensuring that the integral correctly captures the curvature-induced bending of light.

\subsection{Model $\Phi=const.$}
\begin{figure}[ht!]
    \centering
\includegraphics[width=5in,height=5in,keepaspectratio=true]{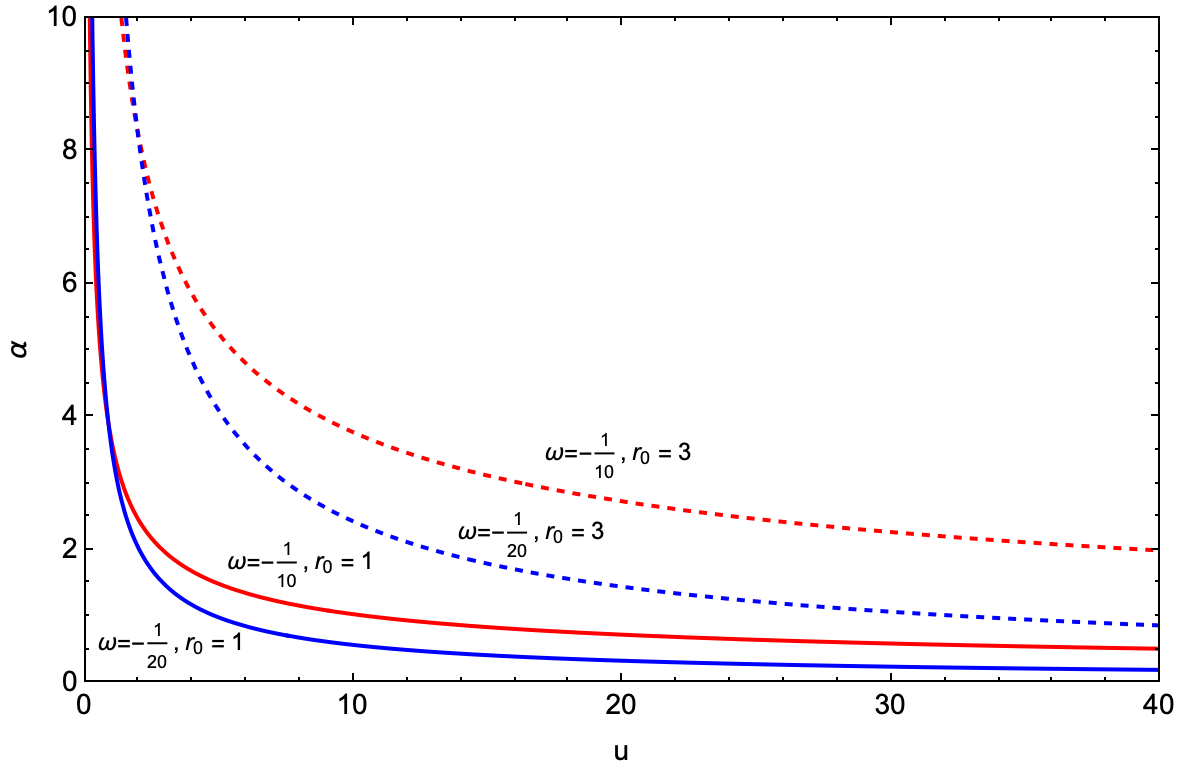}
    \caption{The plots show the deflection angle against the impact parameter $u$ using Eq.(\ref{al1}). The solid-red curve corresponds to $\omega=-1/10$, while the solid-blue curve corresponds to $\omega=-1/20$ and $r_{0} = 1$. The dashed-red curve corresponds to $\omega=-1/10$, while the dashed-blue curve corresponds to $\omega=-1/20$ and $r_{0} = 3$.}
    \label{alpha1}
\end{figure}

In this case, we consider Eq.(\ref{ab}) and Eq.(\ref{RR}) and compture $R$ to yield
\ba
R=-\frac{(3 \omega+1) r_{0}}{r^3}\left(\frac{r_{0}}{r}\right)^{3 \omega}\,.
\ea
Then, the Gaussian curvature $R$ reads
\ba
K=\frac{R}{2}\approx-\frac{(3 \omega+1) r_{0}}{2r^3}\left(\frac{r_{0}}{r}\right)^{3 \omega}\,.
\ea
Approximating this expression in leading order, the deflection angle reads
\ba
\alpha \approx \frac{r_{0} \pi ^{3 \omega +2} \left(\frac{r_{0}}{u}\right)^{3 \omega }}{4 (6 \omega +4) (6 u \omega +u)}\left(r_{0} \pi^{3 \omega } (3 \omega +2) \left(\frac{r_{0}}{u}\right)^{3 \omega }+24 \omega +4\right)\,.\label{al1}
\ea
The deflection angle $\alpha$ strongly depends on the value of $\omega$, which governs the exotic fluid distribution displayed in Fig.(\ref{alpha1}). For negative values of $\omega$, the deflection angle grows more slowly with the impact parameter $u$, indicating weaker lensing effects.
Positive values of $\omega$ lead to stronger lensing since the term 
$(r_{0}/u)^{3\omega}$ grows more slowly with distance, maintaining significant curvature effects even far from the throat. This behavior makes the wormhole resemble traditional gravitational lenses like black holes. A larger throat radius $r_{0}$ increases the deflection angle, meaning that wormholes with wider throats exert a stronger gravitational influence on passing light rays.
As the impact parameter $u$ increases, the deflection angle decreases. This behavior is consistent with standard lensing, where objects further from the source experience less bending of light. For wormholes with smaller throat radii, the lensing effect becomes more localized near the throat.

\subsection{Model $\Phi=r_{0}/r$}
In the second case, we consider Eq.(\ref{ab}) and Eq.(\ref{RR}) and compure $R$ to yield
\ba
R=\frac{r_{0} e^{\frac{2 r_{0}}{r}} (r+r_{0})}{r^5}\left(2 r-\left(\frac{r_{0}}{r}\right)^{3 \omega } (3 r \omega +r+2 r_{0})\right)\,.
\ea
Then, the Gaussian curvature $R$ reads
\ba
K=\frac{r_{0} e^{\frac{2 r_{0}}{r}} (r+r_{0})}{2r^5}\left(2 r-\left(\frac{r_{0}}{r}\right)^{3 \omega } (3 r \omega +r+2 r_{0})\right)\,.
\ea
Approximating this expression in leading order, the deflection angle reads
\ba
\alpha &\approx& -\frac{\pi^2 r_{0}}{24 u^4}\Bigg(-4 u^2 (\pi  r_{0}+3 u)+\nonumber\\&+&\frac{4 u \pi ^{3 \omega } \left(\frac{r_{0}}{u}\right)^{3 \omega } \left(\pi^2 r_{0}^2 (3 \omega +2)+\pi  r_{0} u (3 \omega +2) (3 \omega +4)+3 u^2 (\omega +1) (3 \omega +4)\right)}{(\omega +1) (3 \omega +2) (3 \omega +4)}\nonumber\\&+&\frac{r_{0} \pi^{6 \omega +1} \left(\frac{r_{0}}{u}\right)^{6 \omega } \left(6 \pi^2 r_{0}^2 (2 \omega +1)+3 \pi  r_{0} u (\omega +1) (6 \omega +5)+u^2 (3 \omega +2) (6 \omega +5)\right)}{(2 \omega +1) (3 \omega +2) (6 \omega +5)}\Bigg)\,.\label{al2}
\ea
\begin{figure}[ht!]
    \centering
\includegraphics[width=5in,height=5in,keepaspectratio=true]{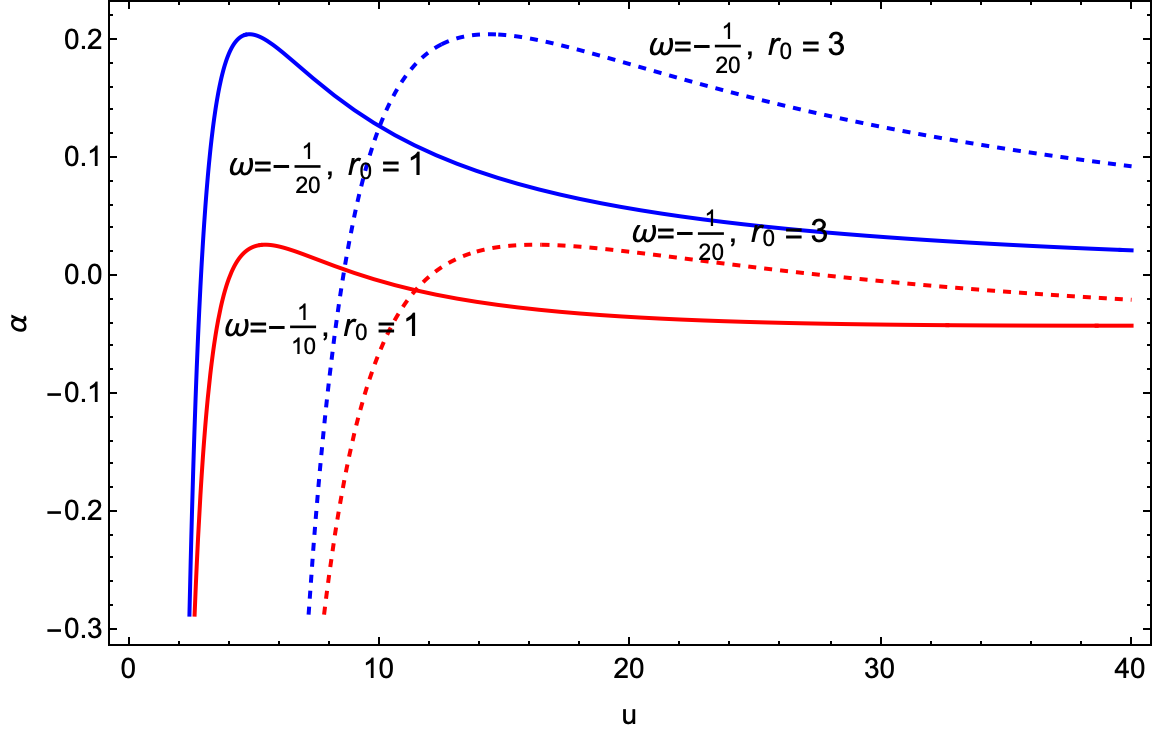}
\caption{The plots show the deflection angle against the impact parameter $u$ using Eq.(\ref{al2}). The solid-red curve corresponds to $\omega=-1/10$, while the solid-blue curve corresponds to $\omega=-1/20$ and $r_{0} = 1$. The dashed-red curve corresponds to $\omega=-1/10$, while the dashed-blue curve corresponds to $\omega=-1/20$ and $r_{0} = 3$.}
    \label{alpha2}
\end{figure}
The deflection angle includes terms of the form $(r_{0}/u)^{3\omega}$ and $(r_{0}/u)^{6\omega}$, reflecting the multi-layered dependence on the ratio between the throat radius $r_{0}$ and the impact parameter $u$ illustrated in Fig.(\ref{alpha2}). For small impact parameters (light passing close to the throat), the higher-order terms dominate, resulting in larger deflection angles. This indicates that the wormhole's gravitational lensing effect is most prominent near its core. For larger values of $u$, these terms decay rapidly, reducing the deflection, which aligns with standard gravitational lensing behavior. When $\omega<0$, the higher-order terms decay more slowly, causing the wormhole to exert its influence over a wider range of impact parameters. This leads to a broader lensing effect, which could be detectable at large distances.

\section{Conclusion}

In conclusion, this study of Kiselev-inspired wormholes has demonstrated that exotic matter distributions can support traversable wormhole geometries. By extending the Kiselev framework, originally developed for black hole solutions, to wormholes, we have shown that only specific anisotropic fluids satisfy the energy conditions necessary to maintain a traversable structure. Our analysis, which examined two distinct redshift functions—constant and inversely dependent on the radial coordinate—revealed that fluids with equation of state parameters between $-1/3$ and $0$ are ideal candidates for sustaining wormholes without extreme violations of energy conditions, thus making them promising models in the context of traversable wormholes. Furthermore, we explored the effects of weak gravitational lensing in these wormhole configurations, showing that gravitational lensing varies based on the choice of redshift function and fluid parameters. The shadows and photon spheres for both static and rotating Kiselev wormholes are worth investigating, see Refs.\cite{Tangphati:2023mpk,Saleem:2024kld}, providing a pathway for potential observational confirmation of Kiselev-inspired wormholes. 

We investigated the effective potential for photons in Kiselev-inspired wormholes, considering two distinct redshift functions: constant and inversely dependent on the radial coordinate. The analysis revealed that the redshift function significantly impacts the effective potential and, consequently, the motion of photons. The results indicate that the photon sphere and associated gravitational lensing effects depend on the wormhole's structural parameters, such as the throat radius and the equation of state parameter. Specifically, variations in the redshift function introduced distinct curvature effects, which influence the gravitational bending of light. Our work also evaluated the amount of exotic matter required to sustain these wormholes using the volume integral quantifier. The results suggest that in certain cases, particularly when the redshift function is constant, the amount of exotic matter can be minimized. This result is advantageous compared to other wormhole models, which often require large and potentially unrealistic quantities of exotic matter.

Another intriguing area of study is the investigation of particle trajectories, which could provide valuable insights into the behavior of particles and the influence of gravitational effects near wormholes (see, for instance, Ref.\cite{Turimov:2022iff}). Additionally, perturbation analyses—such as examining "quasi-normal modes" commonly studied in black hole physics \cite{Volkel:2022khh}—are highly significant and could, in principle, be extended to the study of wormhole physics. Furthermore, an optical observational signature that differentiates asymmetric thin-shell wormholes from black holes has been explored in Ref.\cite{Peng:2021osd}.

Detecting wormhole shadows presents a challenging but ambitious objective that necessitates advanced observational technology and meticulous astrophysical data analysis. Theoretical predictions suggest that the shadow cast by a traversable wormhole exhibits unique characteristics that could distinguish it from black hole shadows. The Event Horizon Telescope (EHT), the most sophisticated instrument currently available for high-resolution imaging of compact objects \cite{EventHorizonTelescope:2019uob,EventHorizonTelescope:2019jan,EventHorizonTelescope:2019ths,EventHorizonTelescope:2019pgp,EventHorizonTelescope:2019ggy}, successfully captured the first-ever direct image of a black hole shadow in $M87^{}$ and subsequently in Sagittarius $A^{}$. Given its capability to resolve structures at scales comparable to a few Schwarzschild radii, the EHT could potentially detect deviations from standard black hole shadows, offering clues about exotic compact objects such as wormholes. Similarly, the Large Synoptic Survey Telescope (LSST) \footnote{http://www.lsst.org/lsst} will provide frequent high-resolution imaging of the sky, which may enable the identification of transient lensing events associated with wormholes. Several critical observational markers could help differentiate wormholes from black holes, including angular resolution, polarization signatures, lensing effects, and variations in shadow structures.

A significant breakthrough was reported in the latest Sloan Digital Sky Survey Quasar Lens Search (SQLS) based on SDSS II, which set the first cosmological constraints on negative-mass compact objects and Ellis wormholes \cite{Takahashi:2013jqa}. In alignment with our current research, we seek to further validate our results by comparing them with observed cosmic distributions of negative-mass compact objects. With the increasing precision of astrophysical observations, future searches for wormholes in astrophysical environments will become more refined and sensitive. However, distinguishing wormholes from black holes based solely on their shadows and photon trajectories remains a substantial challenge. Further advancements in both observational techniques and theoretical models are necessary to confirm their existence, leaving these complex issues open for future exploration.

For future research, several exciting directions can be pursued. One potential avenue is the study of time-dependent wormhole solutions within the Kiselev framework, which could provide insights into the stability and dynamics of such structures in evolving cosmological settings. Additionally, embedding these wormholes in more complex environments, such as those influenced by scalar or electromagnetic fields, may yield more realistic models, making the theoretical predictions more applicable to observations. Investigating the coupling between Kiselev-inspired wormholes and other modified gravity theories could also enrich the current understanding of the interplay between exotic matter and spacetime geometry.

\begin{acknowledgments}
This work is financially supported by Thailand NSRF via PMU-B under grant number PCB37G6600138.
\end{acknowledgments}

\end{document}